\documentclass[11pt]{article}

\usepackage{amsfonts,amscd,amssymb,amsmath,amsthm,latexsym}

\usepackage{xcolor}
\usepackage{graphicx, subfigure,color,float}
\def\bea{\begin{eqnarray}}
\def\eea{\end{eqnarray}}
\def\ba{\begin{array}}
\def\ea{\end{array}}

\def\bi{\begin{itemize}}
\def\ei{\end{itemize}}

\newcommand{\GG}{\mathcal{G}}
\newcommand{\HH}{\mathcal{H}}

\date{}

\begin{document}

\title{A $p$-Adic Matter in a Closed Universe}

\author{\small Branko Dragovich$^{1,2}$ \\
\small $^1$University of Belgrade, Institute of Physics,  11080  Belgrade, Serbia; dragovich@ipb.ac.rs \\
\small $^2$Mathematical Institute of the Serbian Academy of Sciences and Arts, 11000 Belgrade, Serbia }

\maketitle

\abstract{In this paper, we introduce  a new type of  matter that has origin in $p$-adic strings, i.e., strings with a $p$-adic worldsheet.  We investigate  some properties of this $p$-adic matter, in particular its cosmological aspects. We start with crossing symmetric scattering amplitudes for $p$-adic open strings and related effective nonlocal and nonlinear Lagrangian which describes tachyon dynamics at the tree level. Then, we make a slight modification of this Lagrangian and obtain a new Lagrangian for non-tachyonic scalar field. {Using this new Lagrangian in the weak field approximation as a matter in Einstein gravity with the cosmological constant, one obtains an exponentially expanding FLRW closed universe.} At the end, we discuss the obtained results, i.e., computed mass of the scalar $p$-adic particle, estimated radius of related closed universe and noted $p$-adic matter as a possible candidate for dark matter.
 }



\section{Introduction}

$p$-adic numbers were invented (discovered) by mathematician K. Hansel in 1897. Their initial use in physical systems modeling was conducted by I. V. Volovich~\cite{volovich1} in 1987 by construction of some string scattering  amplitudes in terms of $p$-adic analysis; see also~\cite{freund0}.
This work has induced a lot of activity, not only in $p$-adic string theory, but also in many other sectors of modern mathematical and theoretical physics---what is now known as  $p$-{adic mathematical physics}.  We refer to the reviews~\cite{freund1,VVZ,dragovich1, dragovich2}. 

Let us recall that classical theoretical physics uses mathematical methods based on real numbers, whereas quantum theory is inherently related to mathematics with complex numbers, which are algebraic extensions of   real numbers. General relativity combined with quantum mechanics predicts the Planck length as the smallest length that can be measured~\cite{VVZ}. In other words, {there is breakdown of  the Archimedean axiom at the Planck scale and the problem of how to use methods with real and complex numbers emerges, since their geometrical properties are based on the Archimedean norm.} Then, the following question arises: are there some other numbers that could be relevant to approach the very small space--time (Planck) length? {A possible answer could be related to a hypothesis~\cite{volovich2} which assumes  that space--time, at very short  distances, may be non-Archimedean (ultrametric)} and $p$-adic numbers could play some significant role. If so, then string worldsheets may be not only real but also  $p$-adic. This was realized by the construction of some new string amplitudes replacing a real worldsheet by its analog with $p$-adic numbers. Strings with $p$-adic-valued worldsheets are called {\it $p$-adic strings}.

 Progress in $p$-adic string theory has been mainly developed  in two directions, namely, towards the $p$-adic analog of conformal field theory, in particular AdS/CFT correspondence, e.g., \mbox{see~\cite{gubser,marcolli}};  and along an effective  Lagrangian~\cite{freund2,frampton1} for the scalar field, that describes all scattering amplitudes  on the tree level, see~\cite{freund1,VVZ} as a review for initial research. The research work contained in the present paper is based on this effective Lagrangian.

Let us mention some interesting points related to the  effective Lagrangian \eqref{2.3.2} for $p$-adic open strings, which was constructed in 1988~\cite{freund2,frampton1}. Since it does  not contain $p$-adic ingredients, but only real terms,   there is no need to know and use $p$-adic analysis, which simplifies further elaboration of $p$-adic string theory. In fact, this Lagrangian \eqref{2.3.2} is exact at the tree level and contains a scalar field (tachyon) with a nonlocal kinetic term and nonlinear potential. It is also worth mentioning the following: demonstration of tachyon condensation~\cite{sen}, connection to the ordinary bosonic string  in the limit $p \to 1$~\cite{gerasimov}, investigation of dynamics with infinitely many time derivatives~\cite{moeller} and exact solutions~\cite{vladimirov1}, inflation~\cite{barnaby} and some other features (see reviews~\cite{freund1,dragovich1,dragovich2}).

Let us recall that $p$-adic strings  have connections with ordinary strings not only in the limit $p \to 1$~\cite{gerasimov,zuniga1,zuniga2} but also through an adelic product formula of ordinary and $p$-adic crossing symmetric Veneziano amplitudes~\cite{freund3,arefeva1} (see next section
). Despite these connections, $p$-adic strings have been treated as auxiliary constructions with respect to  ordinary strings, often as a toy version of the ordinary ones. However, if ordinary matter has its origin in ordinary strings, {why could $p$-adic strings not generate  $p$-adic matter?} In this paper, we consider how non-tachyonic matter can be obtained from  Lagrangian for $p$-adic tachyons and demonstrate that this new matter makes sense in the case of
a closed universe.

This article is organized as follows:  In Section~\ref{sec2}, some basic facts  about $p$-adic numbers, adeles, amplitudes for scattering of open $p$-adic strings and an effective nonlocal Lagrangian with an equation of motion  are presented. Section~\ref{sec3} is devoted to $p$-adic matter;  the Lagrangian for $p$-adic strings is slightly modified to obtain the well-defined new one, the dynamics of $p$-adic scalar particles is  considered in weak field approximation and a cosmological solution is found and presented in the case of a closed universe fulfilled by  $p$-adic matter with the cosmological constant. Concluding remarks, which contain some discussions on the mass of $p$-adic particles and the radius of the related closed universe,  are the subject of Section~\ref{sec4}. There is also an Appendix
with some details on derivation of the equations of motion in the case of a nonlocal scalar field.

\section{On \boldmath$p$-Adic Strings}\label{sec2}

In this section, we recall basic facts about $p$-adic strings. Since they
have $p$-adic worldsheets, we   start with some mathematical background.

\subsection{$p$-Adic Numbers, Adeles and Their Functions}

For those who are not familiar with $p$-adic numbers, adeles and their functions, here are
some basic introductory facts. To this end, it is useful to start with the field of rational numbers  $\mathbb{Q}$,
since $\mathbb{Q}$ is important from a physical and mathematical point of view. In physics, all numerical results
of measurements are rational numbers. In mathematics, {$\mathbb{Q}$ is an  infinite number field}. With respect to a given prime number $p$, any non-zero rational number $x$ can be presented as $x = \frac{a}{b} p^{\nu}$, where $a, \nu \in \mathbb{Z}$ and$\ b \in \mathbb{N}$; further, $a$  and $b$ are not divisible by $p$. Then, by definition, the $p$-adic norm
(also called $p$-adic absolute value) is  $|x|_p = p^{-\nu}$ and $|0|_p = 0$. One can easily show that $|a|_p \leq 1$, for any  $a \in \mathbb{Z}$ and any prime $p$. From the above definition, it follows the strong triangle inequality $|x + y|_p \leq \text{max}
\{|x|_p, |y|_p\}$, i.e., the $p$-adic norm is an example of ultrametric (non-Archimedean) norm. The $p$-adic distance between $x, y \in \mathbb{Q}$ is $d_p(x,y) = |x-y|_p$. In the same manner, as the field $\mathbb{R}$ of real numbers  obtains from $\mathbb{Q}$ by completion with respect to the real distance $d(x,y) = |x-y|$, so the completion of $\mathbb{Q}$ using a $p$-adic distance gives the field $\mathbb{Q}_p$ of $p$-adic numbers,  for any prime number $p$.

Any non-zero $p$-adic number $0 \neq x \in \mathbb{Q}_p$ has unique representation in the following~form:
\begin{align} \label{2.1}
 x = p^{\nu} \big(x_0 + x_1 p + x_2 p^2 + ...  \big),  \quad \nu \in \mathbb{Z} , \quad x_0 \neq 0, \quad x_n \in
 \{0, 1, ..., p-1\},
\end{align}
where  $x_n$ are digits. For instance, $-1 = p-1 + (p-1)p + (p-1)p^2 + ...$ for any given prime number $p$.

There are mainly two kinds of functions with $p$-adic argument, (i) $p$-adic-valued functions and (ii) complex (real)-valued functions. For example, $p$-adic-valued elementary functions are defined by the same infinite power series as in the real case, but their convergence is subject to the $p$-adic distance. There are three typical complex-valued functions of the $p$-adic argument $x$:
\begin{itemize}
\item multiplicative character: $\pi_p (x) = |x|_p^c, \ \ c\in \mathbb{C}$;
\item additive character: $\chi_p (x) = \exp(2\pi i \{x \}_p)$, where $\{x \}_p$ is a fractional part of $x$;
\item characteristic function: $\Omega (|x|_p) = \begin{cases} 1 \ \ \text{if} \ \  |x|_p \leq 1, \\
0 \ \ \text{if} \ \ |x|_p > 1 .
\end{cases} $
\end{itemize}

There is a well-defined integration of complex-valued functions with the Haar measure, see~\cite{VVZ}. For example, 
$\int_{|x|_p\leq 1} |x|_p^{a-1} = \frac{1- p^{-1}}{1- p^{- a}}$, where  $x$ is a $p$-adic variable and  $a$ is a complex number, with $\Re{a} > 0$.

According to the Ostrowski theorem, real and $p$-adic numbers are all possible numbers that can be obtained by completion
of $\mathbb{Q}$ with respect to any nontrivial norm on $\mathbb{Q}$.  {$\mathbb{Q}$ is a common  subfield of  $\mathbb{R}$ and all $\mathbb{Q}_p$.}

Adeles are a concept that takes together real and $p$-adic numbers. By definition,  an adele is the following infinite sequence:
\begin{align} \label{2.2}
\alpha = (\alpha_\infty, \alpha_2, \alpha_3, ..., \alpha_p , ...),
\end{align}
where $\alpha_\infty \in \mathbb{Q}_\infty \equiv \mathbb{R}$ and, for all but a finite set $\mathcal{P}$ of primes $p$, it must be satisfied that $x_p \in \mathbb{Z}_p \equiv \{ x \in \mathbb{Q}_p : |x|_p \leq 1\}$. $\mathbb{Z}_p$ is called a ring of $p$-adic integers.
The set $A_\mathbb{Q}$ of all adeles over $\mathbb{Q}$ can be defined as
\begin{align} \label{2.3}
A_\mathbb{Q} = \bigcup_{\mathcal{P}} \mathcal{A} (\mathcal{P}) , \quad \mathcal{A} (\mathcal{P}) = \mathbb{R}\times \prod_{p \in \mathcal{P}}  \mathbb{Q}_p  \times \prod_{p \notin \mathcal{P}} \mathbb{Z}_p  .
\end{align}

$A_\mathbb{Q}$ is called an adele ring, since it satisfies, component-wise, addition and multiplication.  
Note that the components of an adele can be  rational numbers; thus, $\mathbb{Q}$ is  naturally embedded in $A_\mathbb{Q}$.
Hence, adeles can be viewed as a generalization of rational numbers that takes simultaneously into consideration all their completions.

There are many useful adelic product formulas  which connect  real and all $p$-adic  constructions  of the same form; e.g., see~\cite{VVZ}.
Some simple cases are:
\begin{itemize}
\item $\pi_\infty (x) \prod_p \pi_p (x)  = |x| \prod_p |x|_p = 1$ , \quad $0\neq x \in \mathbb{Q}$;

\item $\chi_\infty (x) \prod_p \chi_p (x) = e^{-2\pi i x} \prod_p e^{2\pi i \{x \}_p} =1 $,  \quad $x \in \mathbb{Q}$.
\end{itemize}

In the next subsection, we present adelic product formulas for string amplitudes.

Above, some very basic properties of $p$-adic numbers and adeles are presented. For more information, we refer to
books~\cite{VVZ,gelfand,schikhof}.

\subsection{$p$-Adic Open String Amplitudes}

It is worth noticing that string theory started with the Veneziano amplitude. Let us recall that, by definition,  the  crossing symmetric Veneziano amplitude for the scattering of ordinary two-open strings is
\begin{align} \label{2.4}
 A_\infty (a, b) &= g_\infty^2 \,\int_{\mathbb{R}}
|x|_\infty^{a-1}\, |1 -x|_\infty^{b-1}\, d_\infty x  \\
 &= g_\infty^2 \, \frac{\zeta (1 - a)}{\zeta (a)}\, \frac{\zeta (1 -
b)}{\zeta (b)}\, \frac{\zeta (1 - c)}{\zeta (c)} , \quad a + b + c =
1 , \label{2.5}
\end{align}
where   $a, b, c \in \mathbb{C}$  are related to kinematical quantities with a condition, $|\cdot|_\infty$ denotes the usual absolute value and $\zeta$ is the Riemann zeta function.
Then, the analogous $p$-adic  Veneziano amplitude is defined as follows~\cite{freund0}:
\begin{align}  \label{2.6}
 A_p (a, b) &= g_p^2 \, \int_{\mathbb{Q}_p} |x|_p^{a-1}\, |1 -x|_p^{b-1}\, d_p x  \\
&= g_p^2 \, \frac{1 - p^{a - 1}}{1 - p^{-a}}\, \frac{1 - p^{b - 1}}{1 - p^{-b}}\, \frac{1 - p^{c -
1}}{1 - p^{-c}} , \label{2.7}
\end{align}
where   $a, b$ and $c $  are the same quantities as in the above real case. It is obvious that the amplitude in \eqref{2.7} is symmetric under any interchange among
$a, b$ and $c$.
Note that the form of Expressions \eqref{2.4} and \eqref{2.6} is the same and contains analogous ingredients. Only the integration
is different---along a real axis in \eqref{2.4} and over $\mathbb{Q}_p$ in \eqref{2.6}. Regarding the integration of the $p$-adic integral in
\eqref{2.6}, one can see~\cite{freund1,VVZ}. Finally, one can say that the difference between $p$-adic and ordinary strings is in their worldsheets, i.e., $p$-adic and real worldsheets, respectively. Both these kinds of strings are related to tachyons~\cite{freund1}.

Recalling the Euler expression of the Riemann $\zeta$ function and taking the product of \eqref{2.7} over all primes, one obtains
the Freund--Witten formula~\cite{freund3} for the above Veneziano amplitudes.
\begin{align}  \label{2.8}
A (a, b) = A_\infty (a, b) \prod_p A_p (a, b) = g_\infty^2 \,\prod_p g_p^2 = const.
\end{align}

Formula \eqref{2.8} tells us that the amplitudes of the above $p$-adic and ordinary strings are on equal footing, that they may be different faces of an adelic string and that complicate ordinary amplitude with the Riemann zeta function can be expressed as the infinite product of inverse
$p$-adic amplitudes, which are elementary and simpler functions. 

By a similar procedure one can define the amplitudes for $p$-adic closed strings and the corresponding adelic formula also exists;
as a review, see~\cite{freund1}. However, this article is devoted  only to $p$-adic open strings.

\subsection{Effective Field Theory for $p$-Adic Open Strings}

It is very interesting and important that there is an effective field theory model that can reproduce the $p$-adic string amplitudes in \eqref{2.7}.  The corresponding action~\cite{freund2,frampton1} for the scalar field $\varphi(x)$  in D-dimensional Minkowski space is
\begin{equation}
S_p =    \sigma_p \int  \ d^Dx \ \Big( - \frac{1}{2} \varphi\ p^{-\frac{1}{2m^2}\Box}\ \varphi + \frac{1}{p+1}\
\varphi^{p+1}  \Big) , \label{2.3.1}
\end{equation}
where $\sigma_p = \frac{m^D}{g_p^2} \frac{p^2}{p-1}$, $p$ is a  prime number and $\Box =- \frac{\partial^2}{\partial{t}^2} + \sum_{i=1}^{D-1}\frac{\partial^2}{\partial{x^i}^2}$ is the d'Alembert operator ($c= 1$) in $D$-dimensional space--time. Note that Action \eqref{2.3.1}
is invariant (symmetric) under discrete transformation $\varphi \rightarrow - \varphi$ if the prime number $p \geq 3$ and is asymmetric when $p = 2 .$ Field $\varphi$ and mass parameter $m$ can also depend on the prime $p$, but, for simplicity, we omit index $p$. A similar, effective field theory was also constructed for closed $p$-adic strings. This model \eqref{2.3.1} describes not only four-point scattering amplitudes \eqref{2.7}
but also all higher (Koba--Nielsen) ones at the tree-level.

The corresponding Lagrangian
\begin{equation}
\mathcal{L}_p =    \sigma_p \ \Big( - \frac{1}{2} \varphi\ p^{-\frac{1}{2m^2}\Box}\ \varphi + \frac{1}{p+1}\
\varphi^{p+1}  \Big) , \label{2.3.2}
\end{equation}
contains a nonlocal kinetic term with infinitely many space--time derivatives in the form $p^{-\frac{1}{2m^2}\Box}$ and nonlinear potential with $\varphi^{p+1}$ interaction.

The equation of motion (EoM) related to Lagrangian \eqref{2.3.2} is
\begin{align} \label{2.3.3}
p^{-\frac{1}{2m^2} \Box} \varphi \equiv  e^{-\frac{\ln{p}}{2m^2} \Box} \varphi  = \varphi^p .
\end{align}

There are  trivial solutions $\varphi = 0, +1$ for any $p$ as well as $\varphi = -1$, when $p \neq 2$. In the Minkowski space, there is a nontrivial
homogeneous and isotropic  time-dependent solution 
\begin{align} \label{2.3.4}
\varphi(t)  = p^{\frac{1}{2(p-1)}} \exp{\Big(\frac{p -1}{2 p \ln{p}} m^2 t^2    \Big)}
\end{align}
and also an inhomogeneous solution in any spatial direction $x^i$
\begin{align} \label{2.3.5}
\varphi(x^i)  = p^{\frac{1}{2(p-1)}} \exp{\Big(-\frac{p -1}{2 p \ln{p}} m^2 (x^{i})^2    \Big)}.
\end{align}

In D-dimensional space--time, the solution is~\cite{vladimirov1}
\begin{align} \label{2.3.6}
\varphi (x) = p^{\frac{D}{2(p-1)}}\, \exp \Big( - \frac{p-1}{2
 \, p \ln p}\, m^2\, x^2 \Big), \quad x^2 = -t^2 + \sum_{i=1}^{D-1} (x^i)^2 .
 \end{align}

For example, the solution in \eqref{2.3.4} can be obtained employing the identity
\begin{align} \label{2.3.7}
 e^{A \partial_t^2} \ e^{B t^2} = \frac{1}{\sqrt{1- 4 AB}} e^{\frac{B t^2}{1 - 4AB}} , \quad  1-4AB > 0 .
\end{align}

All the above solutions of EoM \eqref{2.3.3} are unstable~\cite{frampton2}.

\begin{figure}[H]
\begin{center}
\includegraphics[scale=.70]{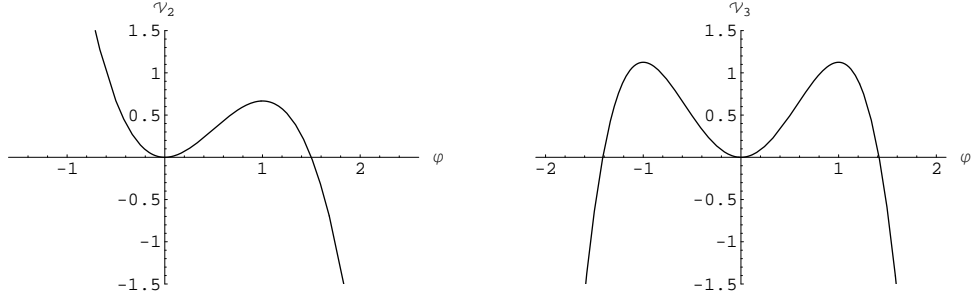}
\caption{The $2$-adic string potential $\mathcal{V}_2 (\varphi)$
(on the \textbf{left}) and  $3$-adic potential $\mathcal{V}_3 (\varphi)$
(on the \textbf{right}) of standard Lagrangian \eqref{2.3.2}, where
potential is presented by Expression \eqref{2.3.8} with
$\sigma_p =1$.}
\end{center}
\label{fig:1}       
\end{figure}

The corresponding  potential $\mathcal{V}_p (\varphi) = -\mathcal{L}_p (\Box = 0)$ of Lagrangian \eqref{2.3.2} is
 \begin{align} \label{2.3.8}
\mathcal{V}_p (\varphi) = \sigma_p \Big[\frac{1}{2}  \varphi^2 - \frac{1}{p +1} \varphi^{p+1}  \Big] ,
 \end{align}
which has local minimum $\mathcal{V}_p (0) =0$ for all $p$ and local maxima $\mathcal{V}_2 (1) = \frac{\sigma_2}{6}$ and
$\mathcal{V}_p (\pm 1) = \sigma_p \frac{p -1}{2(p+1)}$, when $p \neq 2$. When $p=2$ and $p=3$, these potentials are illustrated at Figure~\ref{fig:1}.
When $p\neq 2$, all potentials are an even (symmetric) function of $\varphi$.

Let us consider the above scalar field $\varphi$ in a vicinity of its unstable value $\varphi =1$, i.e., let us take $\varphi = 1 + \eta$, where $|\eta| \ll 1$. Then, EoM \eqref{2.3.3} becomes
\begin{align} \label{2.3.10}
p^{-\frac{1}{2m^2} \Box} (1 + \eta)  = (1 +\eta)^p \approx (1+ p \ \eta),  \quad \Longrightarrow p^{-\big(\frac{M^2}{2m^2} + 1\big)}\ \eta =  \eta ,
\end{align}
which gives $M^2 = - 2m^2$, i.e., the scalar field $\eta \ (\varphi)$ presents a tachyon.

\section{Scalar \boldmath$p$-Adic Matter}  \label{sec3}

We are now going to slightly  modify Lagrangian \eqref{2.3.2} with the intention  to obtain a new Lagrangian for a similar scalar particle which is not a tachyon.
Another direction of research based on Lagrangian \eqref{2.3.2} is towards zeta strings that take into account the effects of $p$-adic strings over all primes $p$; see~\cite{dragovich5} and references therein.

\subsection{Non-Tachyonic $p$-Adic Scalar Field in Minkowski Space}

To this end, for some prime $p$, let us consider  the  transition  $- m^2 \to m^2$ in \eqref{2.3.2}; see an initial consideration in~\cite{dragovich4,dragovich3}. To
differ from a tachyon, we denote this new scalar $p$-adic field by
$\phi.$ Note that, by replacing $- m^2 $ with  $ m^2$, the new related Lagrangian becomes
\begin{align} \label{3.1.1}
 {L}_p (\phi)  = (-1)^{\frac{D}{2}} \, \sigma_p \Big[
-\frac{1}{2}\, \phi \, p^{\frac{\Box}{2 m^2}} \, \phi +
\frac{1}{p+1}\, \phi^{p+1} \Big],
\end{align}
where the change $\sigma_p \to (-1)^{\frac{D}{2}} \sigma_p$ is taken into account.
Depending on space--time dimensionality $D$, we have
\begin{align} \label{3.1.2}
(-1)^{\frac{D}{2}} = \begin{cases}
+1   \quad \text{if} \ \ D= 4k \\
-1   \quad \text{if} \ \ D= 4k +2 \\
+i   \quad \text{if} \ \ D= 4k +1 \\
-i  \quad \text{if} \ \ D= 4k+3 ,
\end{cases}
\end{align}
where $k  \in \mathbb{N}$.  According to \eqref{3.1.2}, it follows that Lagrangian \eqref{3.1.1} can be real only when space--time dimensionality  $D = 2, 4, 6, ...$   Note that the kinetic term is positive when $D = 4k +2$, i.e.,  including $D =
 10$ and  $26$, which are critical dimensions in string theory.

 The equation of motion for the scalar field $\phi$ is
\begin{align}\label{3.1.3}
 p^{\frac{\Box}{2 m^2}}\, \phi = \phi^p
\end{align}
and it has the same trivial solutions as the previous field $\varphi$, i.e., $\phi = 0$ and $\phi =1$ for any $p$ and
 $\phi = -1$, if $p \neq 2$.
There are also nontrivial solutions, such as \eqref{2.3.4}--\eqref{2.3.6}, where one has to replace $m^2$ with $-m^2$.

When $D = 2 + 4k$,  Lagrangian \eqref{3.1.1} is
\begin{align} \label{3.1.4}
 {L}_p (\phi)  =  \sigma_p \Big[
\frac{1}{2}\, \phi \, p^{\frac{\Box}{2 m^2}} \, \phi -
\frac{1}{p+1}\, \phi^{p+1} \Big],
\end{align}
and the corresponding potential becomes
\begin{align} \label{3.1.5}
 {V}_p(\phi) =  \sigma_p \Big[  \frac{1}{p+1}\,
\phi^{p+1} - \frac{1}{2}  \phi^2 \Big].
\end{align}

Note that, now,  $L_p (\phi) = - \mathcal{L}_p (\varphi)$
and   $V_p (\phi) = - \mathcal{V}_p (\varphi)$. 

\begin{figure}[H]
\begin{center}
\includegraphics[scale=.58]{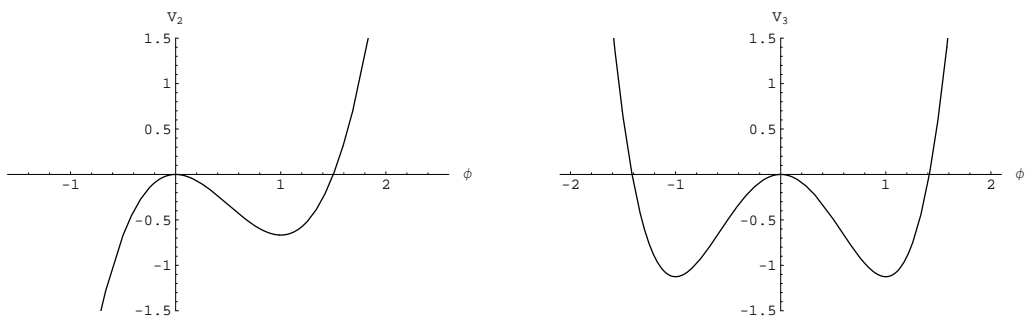}
\caption{New potentials $V_2(\phi)$ and $V_3(\phi)$, which
are defined by \eqref{3.1.5}. This is the same potential as in Figure~\ref{fig:1}, but with opposite  sign. }
\end{center} 
\label{fig:2}       
\end{figure}

The trivial solutions of EoM \eqref{3.1.3} have the following meaning: $\phi_p =0$ is a local maximum and $\phi_2 = +1$ and $\phi_p = \pm 1$, with $\, p\neq 2$, are local minima; see also Figure~\ref{fig:2}.

Let us consider field $\phi$ around minima  $\phi = \pm 1$.  For example,
let $\phi = 1 + \theta$ in the case when $D = 2 + 4 k$. Then, the EoM for weak field
$\theta$, i.e., $|\theta| \ll 1$, becomes
\begin{align}
\label{3.1.6}
 p^{\frac{\Box}{2 m^2}}\, (1 + \theta) = (1+\theta)^p, \quad \Rightarrow
 \quad p^{\frac{\Box}{2 m^2}}\,  \theta =  p \ \theta .
\end{align}

Let us look for a solution of EoM  $p^{\frac{\Box}{2 m^2}}\,  \theta =  p \ \theta$ in some detail. In fact, we have equation 
\begin{align} \label{3.1.7}
e^{\frac{\ln{p}}{2 m^2}\Box}\ \theta = \Big(1 + \frac{\ln{p}}{2 m^2}\Box + \big(\frac{\ln{p}}{2 m^2}\big)^2 \frac{\Box^2}{2!} + ... \Big)\ \theta = p\ \theta,
\end{align}
which has a solution when the following Klein--Gordon equation is satisfied:
\begin{align}
(\Box  - 2m^2) \ \theta = 0 , \quad \text{where} \ \ \ \Box =- \frac{\partial^2}{\partial{t}^2} + \sum_{i=1}^{D-1}\frac{\partial^2}{\partial{x^i}^2}
\end{align}
and we have that $\theta \sim  e^{\pm i(-Et + \vec{k} \vec{x})}$ is a scalar field with  $E^2 = 2 m^2 + \vec{k}^2$.

The above consideration is related to a scalar field in the D-dimensional Minkowski space. In the next
subsection, we want to study some cosmological aspects of   field $\theta$ in 4-dimensional space--time.

\subsection{A Closed Universe with $p$-Adic Matter}

Let us start with a 4-dimensional  gravity with a nonlocal scalar field $\phi$ and cosmological constant $\Lambda$, given by the Einstein--Hilbert action
\begin{equation} \label{3.2.1}
S = \gamma \int \sqrt{-g} \ d^4x \ (R- 2 \Lambda)   + S_m ,
\end{equation}
where $\gamma = \frac{1}{16 \pi G}$, $R$ is the Ricci scalar and
\begin{equation} \label{3.2.2}
S_m = \sigma \int \sqrt{-g} \ d^4x \ \big(\frac{1}{2} \phi F(\Box)\phi  - U(\phi) \big) ,
\end{equation}
where $F(\Box) = \sum_{n=0}^\infty f_n \ \Box^n$ is a nonlocal operator   and $U(\phi)$ is a part of the potential.
Note that, now, $\Box = \nabla_\mu \nabla^\mu = \frac{1}{\sqrt{-g}} \partial_\mu \sqrt{-g} g^{\mu\nu} \partial_\nu$.

According to the variation in Action \eqref{3.2.1},  with respect to $\delta g^{\mu\nu}$ and $\delta \phi$ and the principle of least action
, the equations of motion for gravity field $g_{\mu\nu}$ and scalar field $\phi$ are as~follows:
\begin{align}
&\gamma (G_{\mu\nu} + \Lambda g_{\mu\nu})  - \frac{\sigma}{4} \ g_{\mu\nu} \ \phi F(\Box)\phi + g_{\mu\nu} \ \frac{\sigma}{2}\ U(\phi)
+ \frac{\sigma}{4}\ \Omega_{\mu\nu} (\phi)  = 0, \label{3.2.3}  \\
 &F(\Box)\phi - U'(\phi) = 0 ,  \label{3.2.4}
\end{align}
where
\begin{align}
\Omega_{\mu\nu} (\phi)  =  &\sum_{n=1}^\infty f_n \sum_{\ell=0}^{n-1} \Big[ g_{\mu\nu}\ \big( \nabla^\alpha \Box^\ell \phi
\nabla_\alpha \Box^{n-1-\ell} \phi + \Box^\ell \phi \Box^{n-\ell} \phi \big)   \nonumber \\
&- 2 \nabla_\mu \Box^\ell \phi \nabla_\nu \Box^{n-1-\ell} \phi \Big].  \label{3.2.3a}
\end{align}

For details about the derivation of Equations of motion  \eqref{3.2.3}, we refer to~\cite{dimitrijevic1}; see also Appendix A. 

As a matter of interest, we take the $p$-adic scalar field given by its Action \eqref{3.1.4}
\begin{equation} \label{3.2.5}
S_p = \sigma_p \int \sqrt{-g} \ d^4x \ \Big(\frac{1}{2} \phi\ p^{\frac{1}{2m^2}\Box}\ \phi - \frac{1}{p+1}\
\phi^{p+1}  \Big) ,
\end{equation}
where, again, $\sigma_p = \frac{m_p^D}{g_p^2} \frac{p^2}{p-1}$ and $p$ is a prime number.
Note that, in \eqref{3.2.5}, we have 4-dimensional space--time, but this takes the same signs in the Lagrangian 
as in the case
$D = 4k +2$. The reason for this choice  is to have the correct sign in front of the kinematic term.

The equation of motion  for   $p$-adic field $\phi$ has the same form as the previous one, \eqref{3.1.3},~i.e.,
\begin{align} \label{3.2.6}
p^{\frac{1}{2m^2} \Box} \phi \equiv  e^{\frac{\ln{p}}{2m^2} \Box} \phi  = \phi^p ,
\end{align}
but $\Box$  now depends on the gravity field $g_{\mu\nu}$. It has the same trivial solutions, as in the Minkowski space--time.

The potential $V_p (\phi)$  is already given by Expression \eqref{3.1.5} and $V_2$ and $V_3$ are presented in Figure~\ref{fig:2}.

In the sequel, we are interested in cosmological solutions of the Equations of motion \eqref{3.2.3} and \eqref{3.2.4} in the
homogeneous and isotropic space given by the Friedmann--Lema\^{\i}tre--Robertson--Walker (FLRW) metric, as follows:
\begin{align}\label{3.2.7}
ds^2 = -dt^2 + a^2(t) \Big(\frac{dr^2}{1 -k r^2} + r^2\ d\theta^2  + r^2 \sin^2{\theta}\ d\varphi^2\Big),
\end{align}
where $a(t)$ is the cosmic scale factor and $k=0,+1,-1$ for the plane, closed and open universe, respectively.
Owing to the symmetries of the FLRW metric, there are only two independent Equations of motion \eqref{3.2.3}, which are
usually trace
\begin{align} \label{3.2.8}
 4 \Lambda - R  -\sigma \phi F(\Box) \phi + 2 \sigma U(\phi)  + \frac{\sigma}{4} \Omega = 0
 \end{align}
 and $00$-component
\begin{align} \label{3.2.9}
\gamma (G_{00} - \Lambda)   + \frac{\sigma}{4} \ \phi F(\Box)\phi - \frac{\sigma}{2}\ U(\phi)
+ \frac{\sigma}{4}\ \Omega_{00} (\phi)  = 0,
 \end{align}
where $\Omega = g^{\mu\nu} \Omega_{\mu\nu}$. We   return to \eqref{3.2.8} and \eqref{3.2.9} after some elaboration
of the EoM for field $\phi$ in \eqref{3.2.6}.

Let us look for a solution of \eqref{3.2.6} in a weak field approximation around local minimum $\phi =1$, that is,
$\phi = 1 + \theta$, where $|\theta| \ll 1$. As in \eqref{3.1.6}, again, we have
\begin{align}
\label{3.2.10}
 p^{\frac{\Box}{2 m^2}}\, (1 + \theta) = (1+\theta)^p, \quad \Rightarrow
 \quad p^{\frac{\Box}{2 m^2}}\,  \theta =  p \ \theta ,
\end{align}
where, now,
\begin{align}  \label{3.2.11}
\Box = - \frac{\partial^2}{\partial t^2} - 3 H \frac{\partial}{\partial t} , \quad H = \frac{\dot{a}}{a}
\end{align}
and $H$ is the Hubble parameter.

Equation
\begin{align}\label{3.2.12}
p^{\frac{\Box}{2 m^2}}\,  \theta =  p \ \theta
\end{align}
has solution if there is a solution of $\Box \theta = 2 m^2 \theta$, i.e.,
\begin{align} \label{3.2.13}
 \frac{\partial^2 \theta}{\partial t^2}  + 3 H \frac{\partial \theta}{\partial t} + 2m^2 \theta= 0,
\end{align}
where the Hubble parameter $H =\dot{a}/a$ may be a function of cosmic time, which depends on the scale factor $a(t)$.
The simplest case is $H = constant$ and it corresponds to the scale factor $a(t) = A e^{Ht}$.
When $H$ is constant,   Equation~\eqref{3.2.13} is a linear differential equation with constant coefficients
and has solution in the form $\theta(t) = C \ e^{\lambda t}$, where $\lambda$ must satisfy the quadratic  equation
 \begin{align} \label{3.2.14}
 \lambda^2 + 3 H \lambda + 2 m^2 = 0.
 \end{align}

The solution of Equation~\eqref{3.2.14} has the form $ \lambda_{1,2} = \frac{- 3 H \pm \sqrt{9 H^2 - 8 m^2}}{2}$, where
 $H$ and $m$ can be connected as $H = \pm m$, which yields the simple solutions $\lambda_{+,-} = \pm m .$ It follows that the general solution of Equation~\eqref{3.2.13} can be written
 in the form  
 \begin{align} \label{3.2.15}
  \theta(t) = C_{-}\ e^{-m t} + C_{+}\ e^{+ mt} = \theta_{-}(t)  + \theta_{+}(t),
 \end{align}
 where $C_{-}$ and $C_{+}$ are integration constants. Note that $H$ and $\lambda$ must have opposite sign. Hence, we have the following
 pairs:
 \begin{align}
 &\big(a_{+} (t), \theta_{-} (t)\big): \qquad a_{+} (t) = A_{+} e^{+ mt} , \quad \theta_{-} (t) = C_{-} e^{- mt} ,  \label{3.2.15a}  \\
 &\big(a_{-} (t), \theta_{+} (t)\big): \qquad  a_{-} (t) = A_{-} e^{- mt} , \quad\theta_{+} (t) = C_{+} e^{+ mt}, \quad  .  \label{3.2.15b}
\end{align}

The next step is to explore how the solution in \eqref{3.2.15} satisfies the corresponding equations of motion for a gravitational field. To this end, we have to rewrite the Einstein--Hilbert action with weak field approximation for scalar field $\phi$, i.e.,
we have to rewrite \eqref{3.2.1} in terms of   field $\theta$. The corresponding action is
\begin{align} \label{3.2.16}
S = \gamma \int d^4x \ \sqrt{-g}\ (R - 2 \Lambda) + \sigma_p \int d^4x \ \sqrt{-g}\  \Big(\frac{1}{2}\ \theta p^{\frac{\Box}{2m^2}} \theta - \frac{p}{2} \theta^2 + \alpha_p \Big),
\end{align}
where $\alpha_p = \frac{p-1}{2(p+1)}$.

The potential $V_p (\theta) = - L_p (\Box =0)$ is
\begin{align}\label{3.2.17}
V_p (\theta) = \sigma_p \big(\frac{p-1}{2} \theta^2 -\alpha_p \big)
\end{align}
and it has  the form resembling that of the harmonic oscillator.

We can now return to the Equations of motion \eqref{3.2.8} and \eqref{3.2.9}. With the relevant replacements
\begin{align}\label{3.2.18}
\phi \to \theta, \quad \sigma \to \sigma_p, \quad U(\theta) = \frac{p}{2} \theta^2 -\alpha_p,
\end{align}
we have
\begin{align} \label{3.2.19}
 &\gamma(4 \Lambda - R)  -\sigma_p \ \theta F(\Box) \theta + 2 \sigma_p \ (\frac{p}{2}\ \theta^2 -\alpha_p)  + \frac{\sigma_p}{4}\ \Omega = 0 , \\
 &\gamma(G_{00} - \Lambda)   + \frac{\sigma_p}{4} \ \theta F(\Box)\theta - \frac{\sigma_p}{2}\ (\frac{p}{2} \theta^2 -\alpha_p) + \frac{\sigma_p}{4}\ \Omega_{00} (\theta)  = 0 , \label{3.2.20}
 \end{align}
where
\begin{align} \label{3.2.21}
 F(\Box) = p^{\frac{\Box}{2m^2}}
= \sum_{n=0}^\infty \Big(\frac{\ln p}{2m^2}\Big)^n \frac{1}{n!} \ \Box^n  = \sum_{n=0}^\infty f_n \Box^n .
\end{align}

We show above  that there is a field $\theta$ which satisfies the EoM $p^{\frac{\Box}{2m^2}} \theta = p\ \theta .$ This simplifies the above equations and we come to
\begin{align} \label{3.2.22}
 &\gamma(4 \Lambda - R)    - 2 \sigma_p\alpha_p  + \frac{\sigma_p}{4}\ \Omega (\theta) = 0 , \\
 &\gamma(G_{00} - \Lambda)   + \frac{\sigma_p}{2}\ \alpha_p
+ \frac{\sigma_p}{4}\ \Omega_{00} (\theta)  = 0 . \label{3.2.23}
 \end{align}

Let us recall that, in the FLRW metric,
\begin{align} \label{3.2.24}
 G_{00} = 3  \Big(\frac{\dot{a}^2}{a^2} +\frac{k}{a^2}\Big) , \quad
R = 6  \Big( \frac{\ddot{a}}{a} + \frac{\dot{a}^2}{a^2} +\frac{k}{a^2}\Big) .
\end{align}

Computation \eqref{3.2.24} for  the scale factors $a_{+}(t) = A_{+} e^{+m t}$ and $a_{-}(t) = A_{-} e^{-m t}$ gives  
\begin{align}
&G_{00}^{(+)} = 3 \Big(m^2 + \frac{k}{a_{+}^2}\Big) = 3 \Big(m^2 + \frac{k}{A_{+}^2}\ e^{-2mt}\Big) , \label{3.2.26} \\
&G_{00}^{(-)} = 3 \Big(m^2 + \frac{k}{a_{-}^2}\Big) = 3 \Big(m^2 + \frac{k}{A_{-}^2}\ e^{+2mt}\Big) , \label{3.2.27} \\
&R_{+} = 6 \Big(2m^2 + \frac{k}{a_{+}^2}\Big)  = 6 \Big(2m^2 + \frac{k}{A_{+}^2} \ e^{-2mt} \Big),   \label{3.2.28} \\
&R_{-} = 6 \Big(2m^2 + \frac{k}{a_{-}^2}\Big)  = 6 \Big(2m^2 + \frac{k}{A_{-}^2}\ e^{+2mt} \Big). \label{3.2.29}
\end{align}

Direct calculation of $\Omega (\theta) = g^{\mu\nu}\ \Omega_{\mu\nu}(\theta)$ and $\Omega_{00} (\theta)$,
where (see \eqref{3.2.5})
\begin{align} \nonumber
\Omega_{\mu\nu} (\theta)  =  &\sum_{n=1}^\infty f_n \sum_{\ell=0}^{n-1} \Big[ g_{\mu\nu}\ \big( \nabla^\alpha \Box^\ell \theta
\nabla_\alpha \Box^{n-1-\ell} \theta + \Box^\ell \phi \Box^{n-\ell} \theta \big)\\
&- 2 \nabla_\mu \Box^\ell \theta \nabla_\nu \Box^{n-1-\ell} \theta \Big] ,  \label{3.2.30}
\end{align}
 yields
\begin{align}
&\Omega_{-}  = \Omega (\theta_{-}) = 3 p \ln{p}\ \theta_{-}^2 = 3p \ln{p}\ C_{-}^2 \ e^{-2mt} ,  \label{3.2.31} \\
&\Omega_{+} = \Omega (\theta_{+}) = 3 p \ln{p}\ \theta_{+}^2 = 3 p \ln{p}\ C_{+}^2 \ e^{+2mt} , \label{3.2.32} \\
&\Omega_{00}^{(-)}  = \Omega_{00} (\theta_{-}) = - \frac{3}{2} p \ln{p} \ \theta_{-}^2 = - \frac{3}{2} p \ln{p}  \ C_{-}^2 \ e^{-2mt} , \label{3.2.33} \\
&\Omega_{00}^{(+)} = \Omega_{00} (\theta_{+})  = - \frac{3}{2} p \ln{p} \ \theta_{+}^2 = - \frac{3}{2} p \ln{p} \ C_{+}^2 \ e^{+2mt} . \label{3.2.34}
\end{align}

One can now easily verify that EoM \eqref{3.2.22} and \eqref{3.2.23} are satisfied in the following way:
\begin{align} \label{3.2.35}
 &\gamma(4 \Lambda - R_{\pm})    - 2 \sigma_p\alpha_p  + \frac{\sigma_p}{4}\ \Omega_{\mp} = 0 , \\
 &\gamma(G_{00}^{(\pm)} - \Lambda)   + \frac{\sigma_p}{2}\ \alpha_p
+ \frac{\sigma_p}{4}\ \Omega_{00}^{\mp}   = 0 , \label{3.2.36}
 \end{align}
with the conditions
\begin{align} \label{3.2.37}
6 \gamma m^2 + \sigma_p \alpha_p - 2\gamma \Lambda = 0, \quad  p \ln{p}\ \sigma_p \ A_{\pm}^2 C_{\mp}^2 -8\gamma \ k =0  , \quad k =+ 1 ,
\end{align}
or, in a more explicit form,
\begin{align}
&\Lambda = 3 m^2 + \frac{4\pi G}{g^2} \frac{p^2}{p -1} m^4,  \label{3.2.37a}    \\
&\frac{1}{\big(A_{\pm} C_{\mp} \big)^2} = \frac{2\pi G}{g^2} \frac{p^3 \ln{p}}{p-1}m^4.  \label{3.2.37b}
\end{align}

Therefore, there is a solution of the corresponding equations of motion only in the pair form $\big(a_{\pm}(t), \theta_{\mp}(t)\big).$

\section{Concluding Remarks} \label{sec4}

It is worth noting that \eqref{3.2.37a} contains the connection between the cosmological constant $\Lambda$ and mass $m$ of a $p$-adic scalar particle
(that we   call {\it $p$-adic scalaron} or {\it $p$-scalaron}).
For a small mass $m$, as well as not a big value of prime number $p$ (it  makes  sense  to take $p = 3$) and $g^2 \geq 1$, one can neglect the second term
on the RHS with respect to $3 m^2$. As a result, one obtains $\Lambda \approx 3m^2$, which is written in the natural  units  ($\hbar = c =1$).
In the international system of units (SI), the previous relation should be rewritten as
\begin{align}
 m \approx  \frac{\hbar}{c^2}  \sqrt{\frac{\Lambda}{3}} .   \label{4.1}
\end{align}

From \eqref{4.1}, we can compute the approximate value of mass $m$, where
$$\hbar = 1.05 \times 10^{-34}\,\text{m}^2 \text{kg/s},\ \ c = 3 \times 10^8\,\text{m/s} \  \text{and} \  \Lambda = 3 H^2 \Omega_\Lambda  = 9.8 \times 10^{-36}\,\text{s}^{-2} .$$

We obtain that the mass of $p$-scalaron is 
\begin{equation}
  m \approx 2.1\times 10^{-69} \text{kg}, \label{4.2}
\end{equation}
which  is about a $10^{-39}$ part of the mass of the electron ($m_e = 9.1 \times 10^{-31}\,\text{kg}$). Note that, in the above approximation \eqref{4.1}, the mass of the $p$-adic scalaron does not depend on $p$.

Equality \eqref{3.2.37b} tells us that the product of the radius $R = A_{\pm}$ of the closed universe under consideration and  amplitude $C_{\mp}$ of the $p$-adic scalaron field is a constant that depends on the $p$-scalaron's  mass $m$. Rewriting \eqref{3.2.37b} and using the SI system, we have
\begin{equation}
A_{+} = P_p \ \sqrt{\frac{\hbar^3}{2\pi c}}\ \frac{1}{m^2 \sqrt{G}} ,   \qquad  P_p = \frac{g}{C_{-} }\ \sqrt{\frac{p-1}{p^3 \ln{p}}} .
\end{equation}

Now, one can estimate radius the $R = A_{+}$ of the related closed universe (where $G = 6.67 \times 10^{-11}\,\text{m}^2 \text{kg}^{-1} \text{s}^{-2}$) , that is,
\begin{equation}
R = P_p \ \sqrt{\frac{\hbar^3}{2\pi c}}\ \frac{1}{m^2 \sqrt{G}} \approx P_p \ \times 10^{87} \text{m} ,
\end{equation}
which is a huge number, many times larger than the radius of our observable universe
.

Since a $p$-scalaron has an extremely small  mass \eqref{4.2}, it is unlikely  to be detected in laboratory experiments. However, if the density of $p$-scalarons   is sufficiently large at the galactic scale, then they may play a significant  role as dark matter.
 In addition to gravitational interaction, $p$-scalarons have
also nonlinear and nonlocal self-interaction that gives a solitonic form  to the  effective scalar field  in the $4$-dimensional Minkowski space, i.e.,
\begin{align} \label{4.3}
\varphi (x) = p^{\frac{2}{(p-1)}}\, \exp \Big(  \frac{p-1}{2
 \, p \ln p}\, m^2\, x^2 \Big), \quad x^2 = -t^2 + \sum_{i=1}^{3} (x^i)^2 .
 \end{align}

Note that some  dark matter effects at the cosmic scale can be obtained as nonlocal modification of the Einstein gravity; see~\cite{dimitrijevic2}.
A role of nonlocality  in cosmic dark energy, bouncing and cosmic acceleration is also considered in the framework of string field theory; e.g., see~\cite{arefeva2,arefeva3} and references therein. 

The main results presented in this paper are:
\begin{itemize}
\item Construction of Lagrangian  for $p$-adic matter field and investigation of its equation of motion in weak field approximation.

\item  It is shown that a closed universe fulfilled by $p$-adic matter and a cosmological constant has an  exponential expansion.

\item  A connection between the mass of $p$-adic scalar particle and the cosmological constant is obtained.

\item  The mass of $p$-adic scalar particle is computed.

\item A formula that connects the radius of the closed universe under consideration with the mass of a $p$-adic scalar particle is obtained
and the corresponding radius is estimated.

\item The corresponding notion of $p$-scalaron is proposed and its possible connection with dark matter is conjectured.

\end{itemize}

In the end, it is worth noting how to see some trace of $p$-adic worldsheets in the above effective nonlocal scalar field
theory. To this end, let us consider EoM \eqref{3.1.3} in the simplified form when spatial coordinates are fixed, i.e.,
\begin{align} \label{4.4}
p^{-\frac{1}{2m^2} \frac{\partial^2}{\partial t^2}} \phi (t)   = \phi^p (t).
\end{align}

Using Fourier transform $\phi (t) = \int \tilde{\phi} (E) e^{E t} dE$, one can rewrite  \eqref{4.4} in the form  
\begin{align} \label{4.5}
\int p^{\frac{E^2}{2m^2}} \tilde{\phi} (E)\ e^{E t} dE = \phi^p (t).
\end{align}

Let us recall that, according to the $p$-adic integration with the Haar measure  (e.g., see~\cite{VVZ}), one has
\begin{align} \label{4.6}
\int_{\mathbb{Q}_p\setminus \mathbb{Z}_p} |u|_p^{\alpha -1} \chi_P (u) du = - p^{\alpha -1} , \quad \Re{\alpha} > 0 ,
\end{align}
where $u$ is a $p$-adic variable. Replacing adequately \eqref{4.6} in \eqref{4.5}, one obtains
\begin{align} \label{4.7}
-\int \big(\int_{\mathbb{Q}_p\setminus \mathbb{Z}_p} |u|_p^{\frac{E^2}{2m^2}} \chi_P (u) du   \big) \tilde{\phi} (E)\ e^{E t} dE = \phi^p (t).
\end{align}

A similar procedure can be also conducted in the above effective Lagrangians. One can now conclude that, by some way, a $p$-adic variable $u$ is  related to the $p$-adic worldsheet. Note that the prime number $p$ can be extended to any natural number $n \geq 2$ in Lagrangians \eqref{2.3.2} and \eqref{3.1.4}, but, when $n \neq p$, there is no analogue of Equation~\eqref{4.6} and no direct connection with a $p$-adic string.


\section*{Acknowledgments}

The author wishes to thank Ivan Dimitrijevic, Zoran Rakic and Jelena Stankovic for useful discussions.





\section*{Appendix:\ Derivation of  \boldmath$\Omega_{\mu\nu} (\phi)$}

Here, there are some  main steps  in the derivation of  $\Omega_{\mu\nu} (\phi)$ in EoM  \eqref{3.2.3}, where $\Omega_{\mu\nu} (\phi)$ is given by Equation~\eqref{3.2.3a}; for more details, see~\cite{dimitrijevic1}.

What has to be conducted is the elaboration  of the  variation in the nonlocal operator  $ \delta \mathcal{F} (\Box) = \sum_{n=1}^{+\infty} f_n \ \delta \Box^n .$

Let $\HH$ and $\GG$
 be scalar fields. Then, for any natural number  $n$, we have
\begin{align*}
\delta I &=\int_M \HH \delta (\Box^n \GG) \sqrt{-g} \; d^4x = \int_M \HH \delta (g^{\mu\nu}\nabla_\mu\nabla_\nu  \Box^{n-1}\GG) \sqrt{-g} \; d^4x \\
&= \int_M \HH \left(\nabla_\mu\nabla_\nu\; \Box^{n-1}\GG \ \delta g^{\mu\nu} + g^{\mu\nu} \delta (\nabla_\mu\nabla_\nu  \Box^{n-1}\GG ) \right) \sqrt{-g} \; d^4x \\
&= \int_M \HH \left(\nabla_\mu\nabla_\nu\; \Box^{n-1}\GG \delta g^{\mu\nu} + \Box  \delta (\Box^{n-1}\GG) - (\nabla_\lambda \Box^{n-1}\GG) g^{\mu\nu} \delta \Gamma_{\mu\nu}^\lambda \right) \sqrt{-g} \; d^4x  ,
\end{align*}
where 
\begin{align*}
  g^{\mu\nu} \delta \Gamma_{\mu\nu}^\lambda &= -\frac 12 g^{\mu\nu}\left( g_{\nu\alpha} \nabla_\mu \delta g^{\lambda \alpha} + g_{\mu\alpha} \nabla_\nu \delta g^{\lambda \alpha} - g_{\mu\alpha} g_{\nu\beta} \nabla^\lambda \delta g^{\alpha\beta}\right) \\
  &= -\frac 12 \left( \delta_\alpha^\mu \nabla_\mu \delta g^{\lambda \alpha} + \delta_\alpha^\nu \nabla_\nu \delta g^{\lambda \alpha} - \delta_\alpha^\nu g_{\nu\beta} \nabla^\lambda \delta g^{\alpha\beta}\right) \\
  &= - \frac 12 (2\nabla_\mu \delta g^{\lambda \mu} - g_{\mu\nu} \nabla^\lambda \delta g^{\mu\nu}).
\end{align*}

Further, using   Stokes' theorem, we obtain
\begin{align}
\delta I &=\int_M \HH \Big(\nabla_\mu\nabla_\nu \Box^{n-1}\GG \delta g^{\mu\nu} + \Box  \delta \Box^{n-1}\GG  \nonumber \\
&+\frac 12 \nabla_\lambda \Box^{n-1}\GG (2\nabla_\mu \delta g^{\lambda \mu} - g_{\mu\nu} \nabla^\lambda \delta g^{\mu\nu}) \Big) \sqrt{-g} \; d^4x \nonumber \\
&= \int_M \HH \nabla_\mu\nabla_\nu \Box^{n-1}\GG \delta g^{\mu\nu} \sqrt{-g} \; d^4x + \int_M \HH\;\Box \delta \Box^{n-1}\GG \sqrt{-g} \; d^4x \nonumber \\
&- \int_M \nabla_\mu(\HH \nabla_\lambda \Box^{n-1}\GG)\delta g^{\lambda\mu} \sqrt{-g} \; d^4x \nonumber \\
&+ \frac 12\int_M g_{\mu\nu}\nabla^{\lambda}(\HH \nabla_\lambda \Box^{n-1}\GG) \delta g^{\mu\nu} \sqrt{-g} \; d^4x \nonumber \\
&=  \int_M \HH\; \Box \delta \Box^{n-1}\GG \sqrt{-g} \; d^4x - \int_M \nabla_\mu\HH \nabla_\nu \Box^{n-1}\GG \delta g^{\mu\nu} \sqrt{-g} \; d^4x \nonumber \\
&+ \frac 12\int_M g_{\mu\nu}(\nabla^{\lambda} \HH \nabla_\lambda \Box^{n-1}\GG + \HH \Box \Box^{n-1}\GG) \delta g^{\mu\nu} \sqrt{-g} \; d^4x \nonumber \\
&=  \int_M \HH\; \Box\delta \Box^{n-1}\GG \sqrt{-g} \; d^4x + \frac 12 \int_M S_{\mu\nu}(\HH,\Box^{n-1}\GG) \delta g^{\mu\nu} \sqrt{-g} \; d^4x , \label{A1}
\end{align}
where the following notation is used:
\begin{align}
S_{\mu\nu} (A, B) = g_{\mu\nu} \big(\nabla^\alpha A \ \nabla_\alpha B + A \Box B \big) - 2\nabla_\mu A\ \nabla_\nu B .
\end{align}

The partial integration in the first term of Formula \eqref{A1} yields
\begin{align}
\delta I =  \int_M \Box \HH\; \delta \Box^{n-1}\GG \sqrt{-g} \; d^4x + \frac 12 \int_M S_{\mu\nu}(\HH,\Box^{n-1}\GG) \delta g^{\mu\nu} \sqrt{-g} \; d^4x .
\end{align}

Repeating  the above procedure $n-1$ times, we obtain 
\begin{align}
\delta I=  \int_M \Big(\Box^n \HH\; \delta \GG  + \frac 12\sum_{\ell=0}^{n-1} S_{\mu\nu}(\Box^\ell \HH,\Box^{n-1-\ell}\GG) \delta g^{\mu\nu}\Big) \sqrt{-g} \; d^4x .
\end{align}

Since $\delta \GG = 0$, we have
\begin{align}
\delta I=  \frac 12\sum_{\ell=0}^{n-1} S_{\mu\nu}(\Box^\ell \HH,\Box^{n-1-\ell}\GG) \delta g^{\mu\nu}\Big) \sqrt{-g} \; d^4x .
\end{align}

Finally, taking $\HH= \GG = \phi$,  we obtain
\begingroup\makeatletter\def\f@size{9}\check@mathfonts
\def\maketag@@@#1{\hbox{\m@th\normalsize\normalfont#1}}%
\begin{align}
\Omega_{\mu\nu}(\phi) &= \sum_{n=1}^{\infty} f_n  \sum_{\ell=0}^{n-1} S_{\mu\nu}\big(\Box^\ell \phi,\Box^{n-1-\ell}\phi\big)  \\
&= \sum_{n=1}^\infty f_n \sum_{\ell=0}^{n-1} \Big(g_{\mu\nu}\ \big( \nabla^\alpha \Box^\ell \phi
\nabla_\alpha \Box^{n-1-\ell} \phi + \Box^\ell \phi \Box^{n-\ell} \phi \big)
- 2 \nabla_\mu \Box^\ell \phi \nabla_\nu \Box^{n-1-\ell} \phi \Big) .
\end{align}





\end{document}